\documentclass{PoS}

\title{Double parton distributions of the pion in the NJL model}

\ShortTitle{dPDF of the pion in NJL}

\author{\speaker{Wojciech Broniowski}\thanks{Supported by the Polish National Science Centre (NCN)
Grant 2018/31/B/ST2/01022, the Spanish Ministerio de Economia y
Competitividad and European FEDER funds (grant FIS2017-85053-C2-1-P)
and Junta de Andaluc\'{\i}a grant FQM-225.}\\
        Jan Kochanowski University, 25-406 Kielce, Poland, and \\
       Institute of Nuclear Physics PAN, 31-342 Cracow, Poland \\
        E-mail: \email{Wojciech.Broniowski@ifj.edu.pl}}

\author{Enrique Ruiz Arriola\\
  Departamento de F\'isica At\'omica, Molecular y Nuclear and Instituto Carlos I de F\'{\i}sica Te\'orica y Computacional \\ 
  Universidad de Granada, E-18071 Granada, Spain \\
       E-mail: \email{earriola@ugr.es}}

\abstract{We evaluate the valence double parton distribution (dPDF) of the pion in the Nambu--Jona-Lasinio model.
At the low-energy quark-model scale and in the chiral limit a
particularly simple factorized form $D(x_1,x_2, \vec{q})
= \delta(1-x_1-x_2) F(\vec{q})$ follows, where $x_{1,2}$ denote the longitudinal
momentum fractions of the valence quark and antiquark, and
$\vec{q}$ is their relative transverse momentum.  For
$\vec{q}=\vec{0}$ our result complies to the Gaunt-Sterling sum rules.  
We carry out the necessary dDGLAP evolution to higher scales via the Mellin moments and explore its impact
on the correlation defined as the ratio of dPDF to the product of single parton distributions,
$D(x_1,x_2, \vec{q}=\vec{0})/D(x_1)D(x_2)$. Since the ratios of the valence Mellin moments 
$\langle x_1^n x_2^m \rangle / \langle x_1^n \rangle \langle
x_2^m \rangle $ are invariants of the dDGLAP evolution, they may serve
as robust measures of these correlations. Model predictions, which can be tested in the upcoming lattice
simulations, are provided. We also discuss the transverse form factor related to the dPDF of the pion.}

\FullConference{Light Cone 2019 - QCD on the light cone: from hadrons to heavy ions - LC2019\\
		16-20 September 2019\\
		Ecole Polytechnique, Palaiseau, France}

\begin{document}

More details of this talk and complete references are given
in~\cite{Broniowski:2019rmu}. Double parton distributions (dPDFs) have
recently drawn considerable attention due to evidence for double
parton scattering at the LHC. Model evaluations for the pion, such as
ours or similar studies of~\cite{Courtoy:2019cxq}, are useful in the
light of the awaited lattice QCD results for related quantities, such
as moments of dPDFs~\cite{Zimmermann:2017ctb} or existing two currents
correlators~\cite{Bali:2018nde}. Their simplicity illuminates the
intricate features of the formal definitions of these objects.

\begin{figure}[b]
\begin{center}
\includegraphics[width=0.32\textwidth]{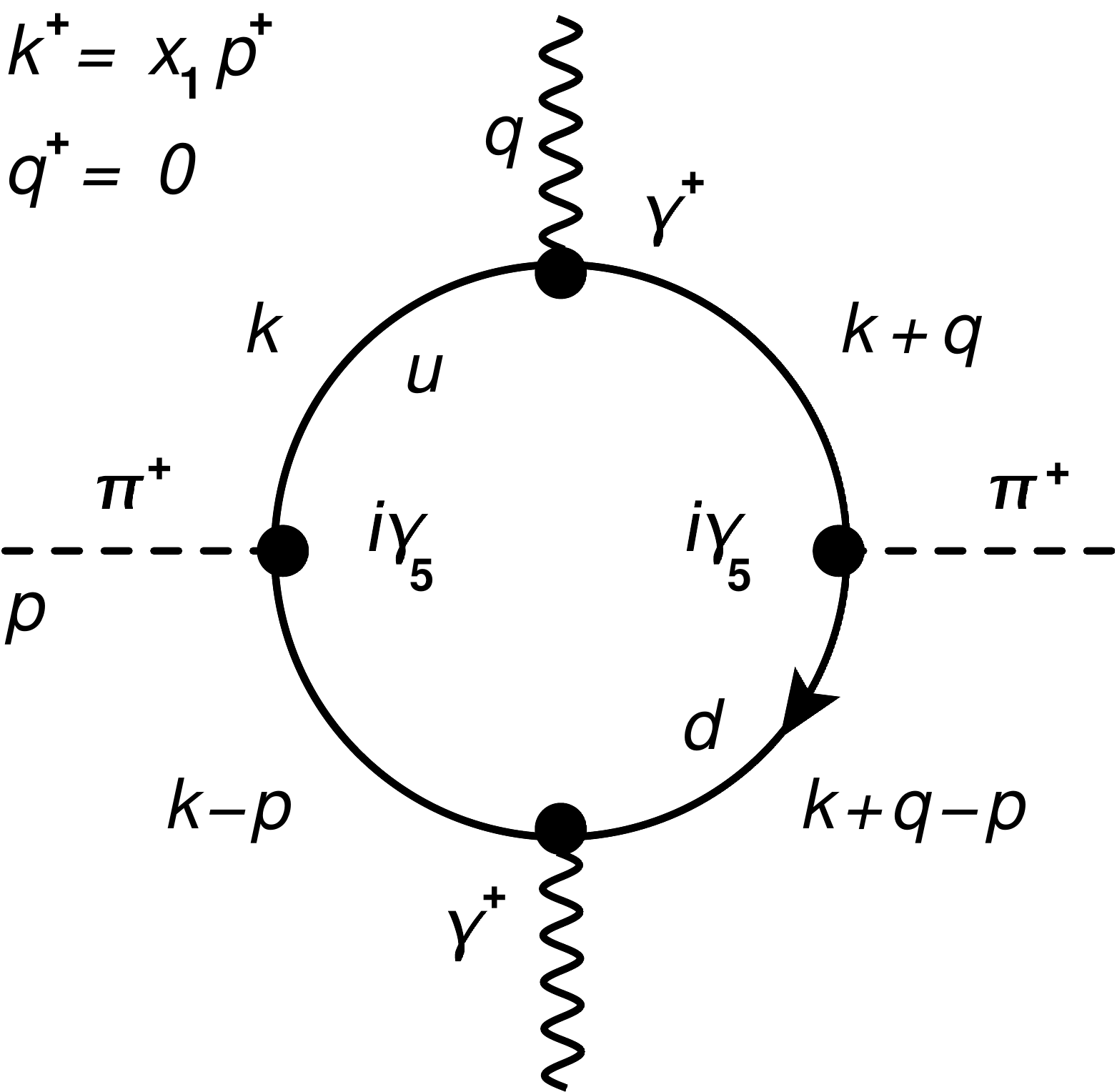} 
\end{center}
\vspace{-5mm}
\caption{Feynman diagram in momentum space for the evaluation of the valence quark dPDF of the pion in chiral quark models.  
A suitable ultraviolet momentum cut-off, implemented in a gauge-invariant way 
via spectral regularization~\cite{RuizArriola:2003bs} or Pauli-Villars method, is understood. \label{fig:diag}}
\end{figure}

The {\em spin-averaged} dPDF~(see \cite{Diehl:2011yj} and references therein) for the parton species $j_{1,2}$ in a hadron with momentum $p$
is defined via the following forward matrix element,
\begin{eqnarray}
D_{j_{1} j_{2}}(x_1,x_2,\vec{y})
 \!\!= \!\! 2 p^+ \int d y^-\,
        \frac{d z^-_1}{2\pi}\, \frac{d z^-_2}{2\pi}\;
          e^{i ( x_1^{} z_1^- + x_2^{} z_2^-) p_{}^+} 
    \langle p |\, {\mathcal{O}_{j_1}(y,z_1)\, \mathcal{O}_{j_2}(0,z_2)}
    \,| p \rangle
    \bigl|_{z_1^+ = z_2^+ = y_{\phantom{1}}^+ = 0\,,
    \vec{z}_1^{} = \vec{z}_2^{} = \vec{0}},  \nonumber \\ \label{eq:defsd}
\end{eqnarray}
where {$\vec{y}$ is the
transverse distance between the two partons and the light-cone
coordinates are \mbox{$a^\pm = (a^0 \pm a^3) /\sqrt{2}$}. 
For the case of quarks and antiquarks considered here, the bilocal color-singlet operators are explicitly
\begin{eqnarray}
\mathcal{O}_{q}(y, z) = \frac{1}{2}\, \bar{q} ( y
- \frac{z}{2}) \gamma^+ q ( y
+ \frac{z}{2} ),  \; \;\; \mathcal{O}_{\bar{q}}(y, z) =
-\frac{1}{2}\, \bar{q} ( y + \frac{z}{2}) \gamma^+  q ( y - \frac{z}{2}). 
\end{eqnarray} 
The summation over color is implicit, whereas the
flavor indices are skipped.  Since the quark and antiquark
coordinates within the operators $\mathcal{O}_{j}$ in
Eq.~(\ref{eq:defsd}) are not split in the transverse direction, the
light-front gauge along straight line paths sets the
Wilson gauge link operators to identity. 

Evaluation according to the diagram of Fig.~\ref{fig:diag} in the chiral limit yields the result of an appealing
simplicity, 
\begin{eqnarray}
D(x_1,x_2, \vec{q})= \delta(1-x_1-x_2) F(\vec{q}), \label{eq:res}
\end{eqnarray}
holding at a low-energy quark model scale. Moreover, it satisfies
the Gaunt-Stirling sum rules~\cite{Gaunt:2009re}, which is nontrivial
to accomplish in parametrization
approaches~\cite{Broniowski:2013xba,Golec-Biernat:2015aza,Broniowski:2016trx}. We note a
factorization of the longitudinal and transverse degrees of freedom,
similarly to the case of single parton distributions (sPDFs) of the
pion in the NJL model. The result of Eq.~(\ref{eq:res}) befits a
simple phase space calculation with $x$-independent matrix
elements~\cite{RuizArriola:1999hk}.

\iffalse
\begin{figure}
\begin{center}
\includegraphics[width=0.39\textwidth]{cpl08.pdf}  \includegraphics[width=0.39\textwidth]{cpl2.pdf} \\
\includegraphics[width=0.39\textwidth]{cplp1000.pdf} \includegraphics[width=0.39\textwidth]{cplp1000000000000.pdf} 
\end{center}
\caption{$D^{\pi^+}_{u \bar{d}}(x_1,x_2)$ evolved with LO dDGLAP equations. \label{fig:D12}}
\end{figure}
\fi

\begin{figure}[t]
\begin{center}
\includegraphics[width=0.39\textwidth]{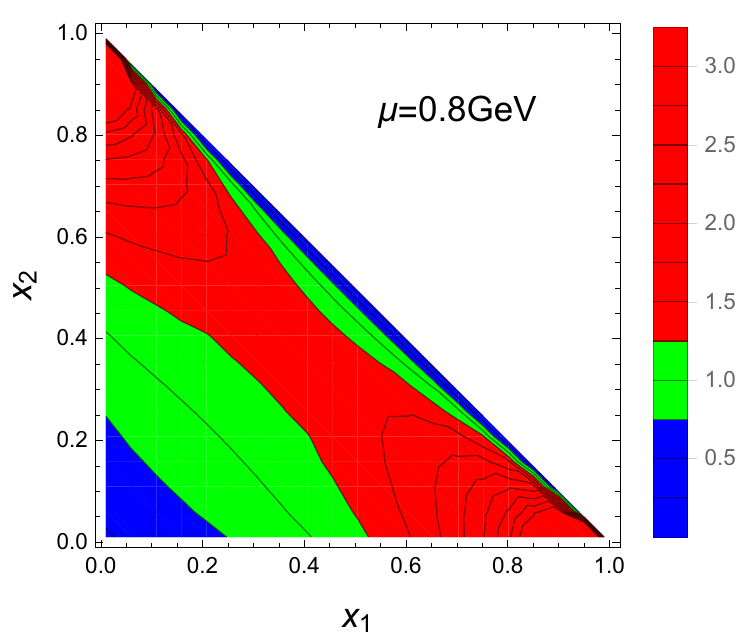}  \includegraphics[width=0.39\textwidth]{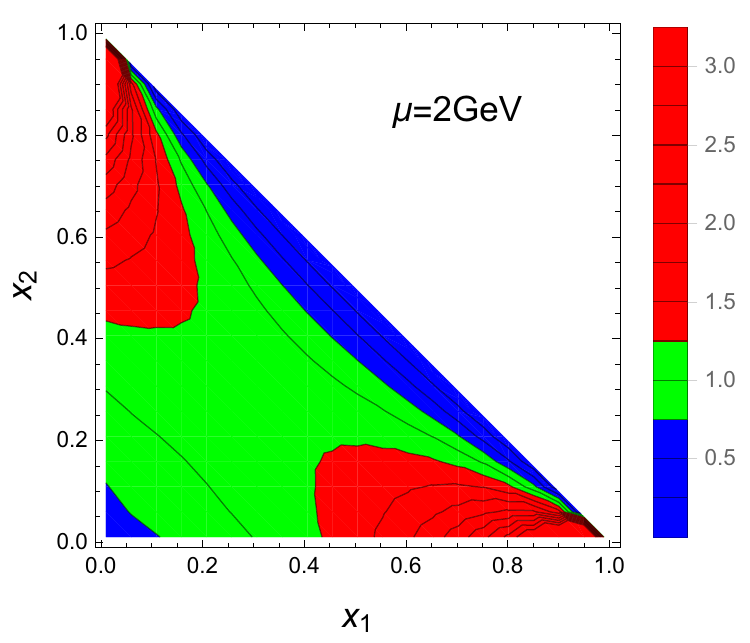} \\
\includegraphics[width=0.39\textwidth]{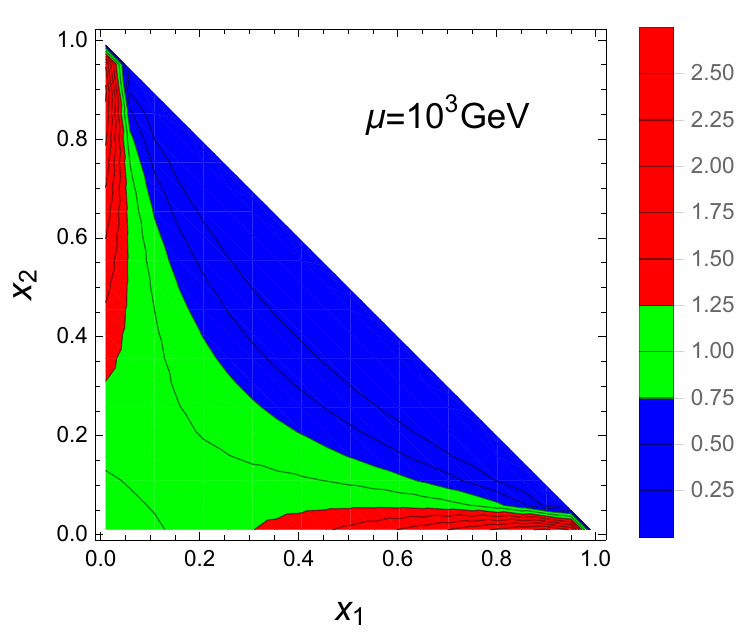}  \includegraphics[width=0.39\textwidth]{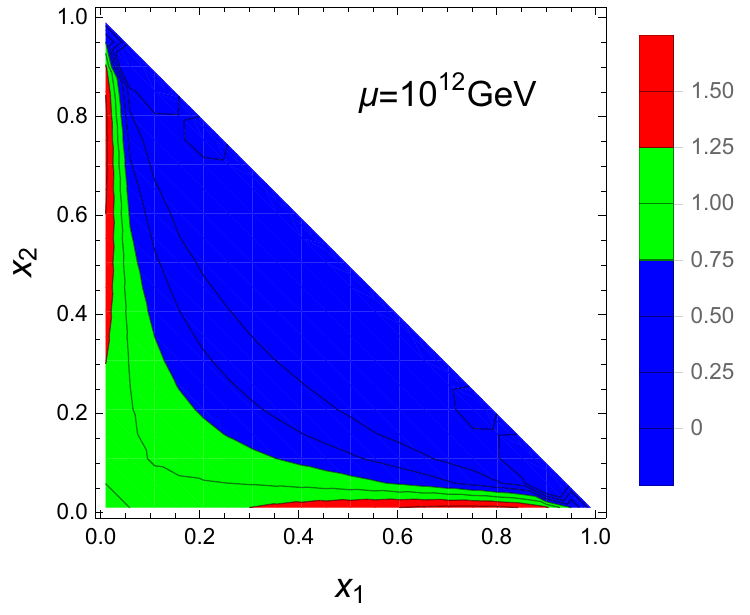}  
\end{center}
\vspace{-7mm}
\caption{Correlation ${D^{\pi^+}_{u \bar{d}}(x_1,x_2, \vec{q}=\vec{0})}/[{D_u(x_1)D_{\bar{d}}(x_2)}]$ 
at increasing evolution scales $\mu$. \label{fig:cor12}}
\end{figure}

QCD evolution is a necessary ingredient of the phenomenology built on
low-energy models, as it allows to bring the results to the energy
domain accessible in experiments and lattice simulations.  The dDGLAP
evolution of the valence Mellin moments is particularly simple, due to
the absence of the inhomogeneous term. It immediately follows that the
ratios $r_{nm}=\langle x_1^n x_2^m \rangle /\langle x_1^n \rangle
\langle x_2^m \rangle$ for the valence distributions are scale
invariant, because the evolution ratios cancel out.  The valence Mellin
moments of sPDFs and dPDFs evolve as
\begin{eqnarray}
&& M_n(\mu)=\left (\frac{\alpha(\mu)}{\alpha(\mu_0)}\right )^{\gamma_n/2\beta_0}M_n(\mu_0), \; \;\;
M_{nm}(\mu)=\left (\frac{\alpha(\mu)}{\alpha(\mu_0)}\right )^{\gamma_{n}/2\beta_0 + \gamma_{m}/2\beta_0 }M_{nm}(\mu_0), \label{eq:anom}
\end{eqnarray}
with $\gamma_i$ denoting the appropriate anomalous dimensions, which
leads to the advocated invariance of $r_{nm}$.  This very simple
feature has far-reaching consequences, as it brings an insight into
the non-perturbative dynamics at low energy scales from the data at high energy
scales.  In the NJL model we find from Eq.~(\ref{eq:res})
$r_{nm}= {(1+n)!(1+m)!}/{(1+n+m)!}$.
The largest ratio is for $m=n=1$. The particular set of values for the moments is specific 
to a given low-energy model. The LO dDGLAP evolution of the lowest Mellin moments of the valence dPDF of
the pion is shown in Fig.~\ref{fig:mom-evol}. Of course, the fall-off
with the scale $\mu$ is controlled by the appropriate anomalous
dimensions in Eq.~(\ref{eq:anom}).

The dPDF form factor in the transverse variable both
in momentum space as well as in coordinate space is shown in
Fig.~\ref{fig:ff} for both the NJL model, PV
regularized~\cite{RuizArriola:2002wr} and the Spectral Quark Model
(SQM)~\cite{RuizArriola:2003bs}. The form factor can be shown~\cite{Broniowski:2019rmu} to mathematically
correspond to a convolution of a wave function with itself. A
remarkable feature seen from Fig.~\ref{fig:ff} is the lack of positivity, which is
surprising but does not contradict any general requirement. For
instance, a convolution of functions with nodes with itself is not necessarily
positive. Future lattice QCD calculations may shed some light on this
interesting problem.

\begin{figure}[t]
\begin{center}
\includegraphics[width=0.45\textwidth]{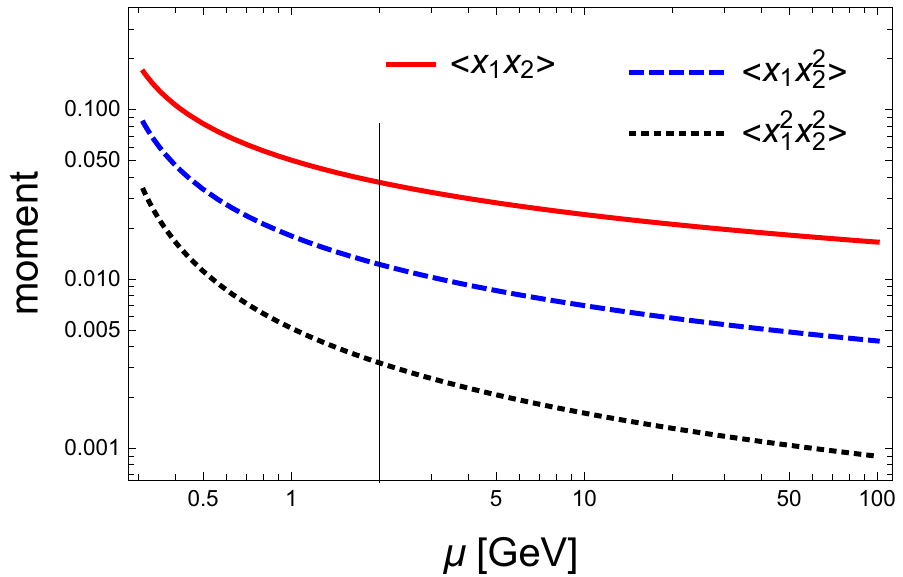}
\caption{Lowest Mellin moments $M_{nm}$ of the valence dPDF of the pion plotted as functions of the evolution scale $\mu$. 
\label{fig:mom-evol}} 
\end{center}
\end{figure}

\begin{figure}[b]
\begin{center}
  \includegraphics[width=0.45\textwidth]{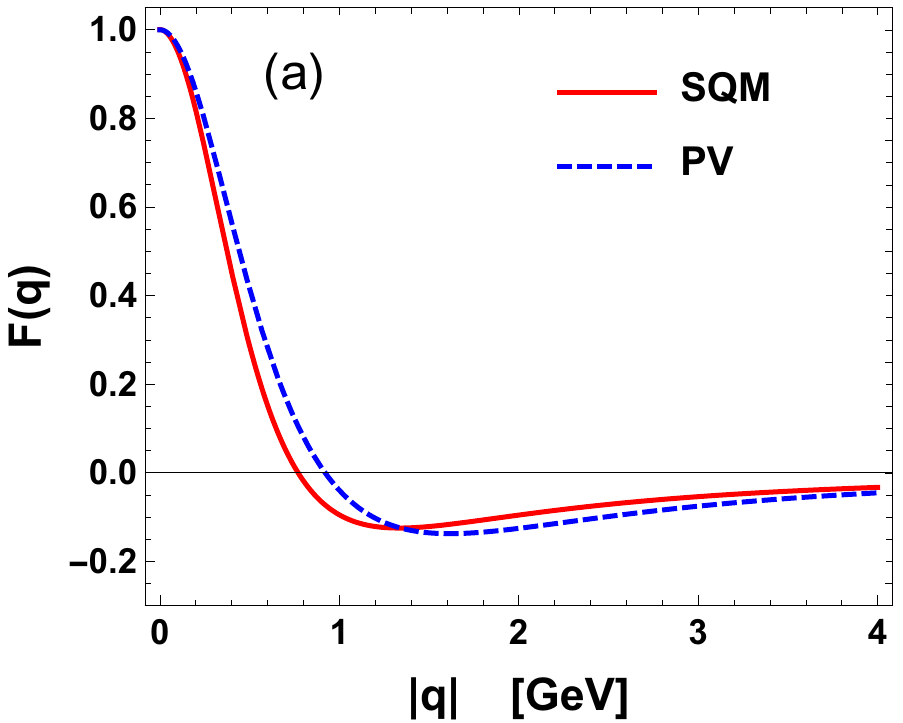}
  \includegraphics[width=0.45\textwidth]{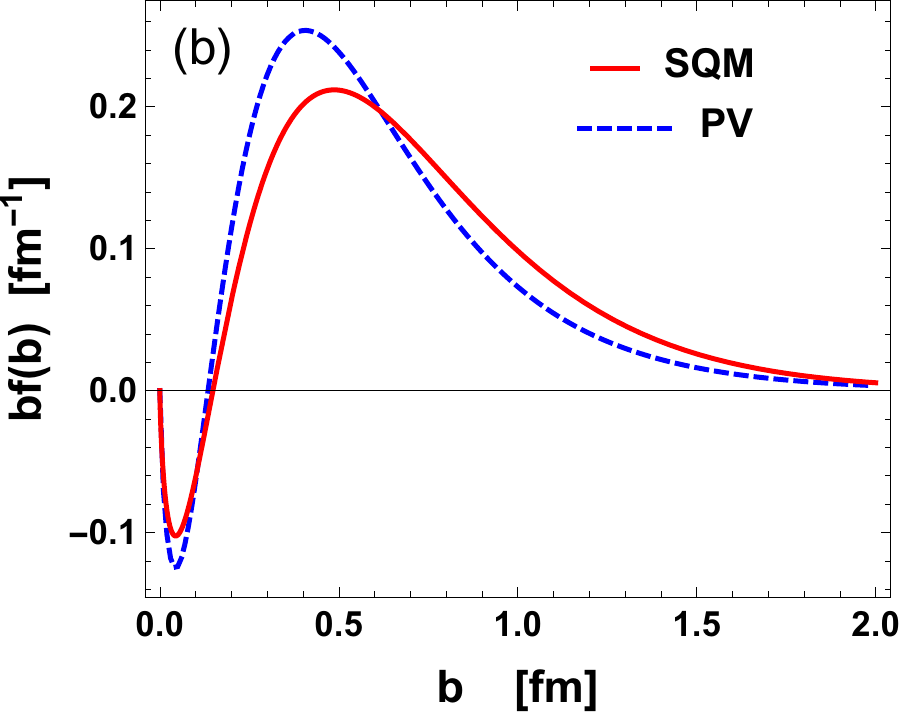} 
\caption{dPDF form factor in the momentum space, $F(\vec q)$, (left panel)
  and coordinate space, $b f(b)$, (right panel), plotted as a functions of the
  corresponding transverse variable for the NJL model with the PV
  regularization and for SQM.  The form factors $F(\vec q)$ and
  $f(\vec b)$ are connected by a Fourier-Bessel transform.
\label{fig:ff}} 
\end{center}
\end{figure}

Finally, we comment on possible lattice determinations of quantities related to
dPDFs. Firstly, on the lattice the currents are local, with $z_1^-$ and $z_2^-$ in Eq.~(\ref{eq:defsd}) set
to zero, which corresponds to the integration over $x_1$ and
$x_2$. Second, the relative time difference of the two currents
in~\cite{Bali:2018nde,Zimmermann:2017ctb} is set to zero, whereas in
Eq.~(\ref{eq:defsd}) unconstrained integration over $y^-$ is carried out. Thus, our
dPDF results for the form factors cannot be directly compared
to~\cite{Bali:2018nde,Zimmermann:2017ctb}, even if the pion mass were
taken to its lattice value of $\sim 300$MeV.

\bigskip

To summarize our main results:
\begin{itemize}

 \item The chiral quark models yields in the chiral limit a factorization of the
 longitudinal and transverse dynamics. At the 
 low-energy quark model scale, the longitudinal distribution is just
 $\delta(1-x_1-x_2)$, as follows from the momentum conservation.
 
 \item We have carried out the LO dDGLAP evolution via the Mellin moments.
 
 \item The Gaunt-Stirling sum rules are explicitly satisfied in our approach.
 
 \item The longitudinal correlation function, defined as the ratio of dPDF
 from the product of two sPDFs, allows for a test
 of the validity of the commonly used product ansatz. The
 conclusion is that at high $\mu$ and small momentum $x_1$ and $x_2$ it is justified 
 within the present experimental kinematics and accuracy.
 
 \item The $r_{nm}$ moments, which are invariants of the evolution,
   are particularly useful objects for future lattice studies, as they
   allow to look down into low energies and scrutinize
   non-perturbative aspects of the pion structure.

\item The dPDF form factor becomes negative above large transverse
  momenta $\sim 0.8$~GeV or below  transverse separations $\sim 0.15$~fm.
  
  %Lattice calculations may allow to discern if this is a true effect.
 
\end{itemize}

\bibliographystyle{JHEP}
\bibliography{dPDF}

\end{document}